
\def\halfspace{\baselineskip=1.5\normalbaselineskip}
\def\Rf{\normalbaselines\parindent=0pt \medskip\hangindent=3pc \hangafter=1 }
\def\AB{\bigskip\parindent=40pt
        \centerline{\bf ABSTRACT}\medskip\halfspace\narrower}
\def\doublespace{\baselineskip=2\normalbaselineskip}
\def\AE{\bigskip\nonarrower\doublespace}
\def\nonarrower{\advance\leftskip by-\parindent
	\advance\rightskip by-\parindent}

\def\GR{{\rm I\kern-.23em R}}
\def\GH{{\rm I\kern-.23em H}}
\def\GC{{\rm\kern.24em
  \vrule width.02em height 1.4ex depth-.05ex \kern-.30em C}}
\def\GF{{\rm I\kern-.23em F}}
\def\GN{{\rm I\kern-.21em N}}
\def\GQ{{\rm\kern.24em \vrule width.02em height1.4ex
depth-.05ex
\kern-.30emQ}}
\def\GO{{\rm\kern.24em \vrule width.02em height1.4ex depth-.05ex
\kern-.30em O}}
\def\sgh{{\rm I\kern-.22em H}}
\def\sgc{{\rm I\kern-.32em C}}
\catcode`@=11
\def\eqalignno#1{\displ@y \tabskip\centering
  \halign to\displaywidth{\hfil$\displaystyle{##}$\tabskip\z@skip
    &$\displaystyle{{}##}$\hfil\tabskip\centering
    &\llap{\rm ##}\tabskip\z@skip\crcr
    #1\crcr}}
\catcode`@=12

\tolerance=500000
\hoffset=.125in
\voffset=.0625in

{\nopagenumbers
\rightline{IASSNS-HEP-93/46}
\rightline{August 1993~~~~~~}

\bigskip
\centerline{{ Proof of Jacobi identity in
generalized quantum dynamics}\footnote{$^*$}{
Submitted to
Nuclear Physics {B}}}
\bigskip\medskip
\centerline{ Stephen L. Adler}
\medskip
\centerline{ Institute for Advanced Study}
\centerline{ Princeton, NJ 08540}
\bigskip
\centerline{ Gyan V. Bhanot}
\medskip
\centerline{ Thinking Machines Corporation}
\centerline{ 245 First Street, Cambridge, MA 02142}
\medskip
\centerline{ and}
\medskip
\centerline{ Institute for Advanced Study}
\centerline{ Princeton, NJ 08540}
\bigskip
\centerline{ John D. Weckel}
\medskip
\centerline{ Physics Department, Princeton University}
\centerline{ Princeton, NJ 08540}
\bigskip
\AB
\baselineskip=20pt
\parindent=.5in
We prove that the Jacobi identity for the generalized Poisson bracket
is satisfied in the generalization
of Heisenberg picture quantum mechanics recently proposed by one of us
(SLA).  The identity holds for any combination of fermionic and
bosonic fields, and requires no assumptions about their mutual
commutativity.
\AE

\vfill\eject}

\pageno=2
\baselineskip=25pt
\centerline{\bf 1. Introduction}

In a recent paper, one of us (SLA) proposed a new quantum dynamics by
generalizing the fundamental field equations from c--number to unitary
or bi--unitary operator gauge invariance [1].  The approach was motivated by an
investigation of quaternionic quantum mechanics and field theory [2],
and more generally, gives a symplectic dynamics for general
non--commutative degrees of freedom, provided only that multiplication
is associative and there exists a
trace permitting cyclic permutation of the non--commuting variables.

An important ingredient missing from the formalism of [1] was the
proof of the Jacobi identity for the generalized Poisson bracket, of
arbitrary polynomial trace functionals of
fermionic and bosonic quantum variables.  The Jacobi identity is
necessary for the correct incorporation of symmetries.  Specifically, if
${\bf A}$ and ${\bf B}$ are conserved symmetry generators, then the
Jacobi identity implies that their generalized Poisson bracket is also a
conserved symmetry generator.  More generally, because of the Jacobi
identity, we expect total trace symmetry generators, such as the
Poincar\'e  generators, to obey a Lie algebra under the generalized Poisson
bracket operation, which is isomorphic to the Lie algebra under
commutation obeyed by the corresponding abstract generators of
symmetries of the total trace Lagrangian.

In this paper we prove that the Jacobi identity
is indeed satisfied for the formalism of [1].  To keep the paper
self--contained, we define below the ingredients of the formalism
necessary for our proof.  The interested reader should consult [1]
and [2] for further details.
\bigskip
\centerline{\bf 2.  The proof}

One starts by defining a Hilbert space $V_H$ (based either on
complex number or quaternionic scalars) which is the direct sum of a bosonic
space
$V_H^+$ and a fermionic space $V_H^-$.  Next, following Witten [3], one
defines an operator $(-1)^F$ with eigenvalue $+1$ for states in
$V_H^+$ and $-1$ for states in $V_H^-$.
Finally, one needs a trace operation ${\bf Tr}\,{\cal O}$ for a
general operator ${\cal O}$, defined by
 $$
{\bf Tr}~{\cal O}= {\rm
Re}\,Tr\,(-1)^F{\cal O}= {\rm Re} \sum_n~\langle n|(-1)^F{\cal
O}|n\rangle.
\eqno\rm (1)
$$
It is easy to show that the trace ${\bf Tr}$ is non--vanishing only for
operators ${\cal O}$ which commute with $(-1)^F$.

Let $\{q_r(t)\}$ be a set of time--dependent quantum variables, which
act as operators on the underlying Hilbert space, with each individual
$q_r$ of either bosonic or fermionic type, defined respectively as
commuting, or anti--commuting with $(-1)^F$.  No other {\it a priori}
assumptions about commutativity of the $q_r$ are made.  The
Lagrangian ${\bf L}[\{q_r\},\{\dot q_r\}]$ is then defined as the trace of a
polynomial function of $\{q_r(t)\}$ and its time derivative $\{\dot
q_r(t)\}$, or as a suitable limit of such functions.  The action
${\bf S}$ is defined as the time integral
of ${\bf L}$, and generalizations of the Euler--Lagrange equations
follow from the requirement that $\delta {\bf S} = 0$ for arbitrary
(same--type) variations of the operators.  Derivatives of ${\bf L}$ with
respect to $q_r$ and $\dot q_r$ are defined by writing the variation
of ${\bf L}$, for infinitesimal variations in the $\{q_r\}$, in the form,
$$
\delta {\bf L} = {\bf Tr}~\sum_r~\left({\delta{\bf L}\over\delta q_r}~\delta
q_r
+{\delta{\bf L}\over\delta\dot q_r}~\delta\dot q_r\right)~,
\eqno\rm (2)
$$
where cyclic permutations of operators inside ${\bf Tr}$ have been used
to order $\delta q_r$ and $\delta\dot q_r$ to the right.  The momentum
$p_r$ conjugate to $q_r$ is defined by
$$
{{\delta {\bf L}}\over{\delta \dot q_r}}=p_r~,
\eqno\rm (3)
$$
and the Hamiltonian ${\bf H}$ is given by
$$
{\bf H}={\bf Tr}~\sum_r~p_r\dot q_r-{\bf L}~.
\eqno\rm (4)
$$

The generalized Poisson bracket of two trace functionals ${\bf
A}[\{q_r\},\{p_r\}]$ and ${\bf B}[\{q_r\},\{p_r\}]$ is then defined as
$$
{\bf \{A,B\}}\equiv{\bf Tr}\sum_r~\epsilon_r\left({\delta{\bf A}\over\delta
q_r}~{\delta{\bf B}\over\delta p_r}-{\delta{\bf B}\over\delta
q_r}~{\delta{\bf A}\over\delta p_r}\right)~,
\eqno\rm (5)
$$
with $\epsilon_r = +1 $ if $q_r$ is bosonic, and $\epsilon_r=-1$ if $q_r$
is fermionic.

The purpose of this paper is to prove that for the formalism defined above,
the Jacobi identity is satisfied.
That is, if ${\bf A}[\{q_r\},\{p_r\}],~{\bf
B}[\{q_r\},\{p_r\}]$, and ${\bf C}[\{q_r\},\{p_r\}]$ are arbitrary polynomial
trace functionals of the operator arguments $\{q_r\}, \{p_r\}$, then,
$$
{\bf [A,B,C]}\equiv {\bf\{A,\{B,C\}\}}+{\bf \{B,\{C,A\}\}}+{\bf
\{C,\{A,B\}\}}~= 0.
\eqno\rm (6)
$$

We first attempted to check or find a counter--example to Eq.~(6) by
generating many computer examples, and using a computer algorithm based
on the manipulation of strings of integer labels to test the Jacobi
identity.  After many successful verifications on the computer, we
found the analytic proof which we present below.

For ease of exposition, we will use a more compact notation.
Derivatives with respect to $q_r$ or $p_r$ of a trace functional ${\bf
A}$ will be denoted by ${\bf A}_r$ and ${\bf A}^r$ respectively.  The
operation ${\bf Tr}$ will be implied by the parentheses $(~)$
and, for the time being, it will be assumed that repeated indices are
summed.

In this notation, the Poisson bracket is given by,
$$
{\bf \{A,B\}}= \epsilon_r\left( {\bf A}_r {\bf B}^r -
                 {\bf B}_r {\bf A}^r \right).
\eqno\rm (7)
$$

It is useful to illustrate with an example how derivatives are computed.
Consider the case where we have two kinds of field variables
$q_1,p_1$ and $q_2,p_2$.  Given the trace functional
${\bf A} =  \left( q_1 p_1 q_2 q_1 p_2 q_1 \right)$, its
derivative with respect to $q_1$ is denoted by
${\bf A}_1$ and is given by
$$
{\bf A}_1 = q_1 p_1 q_2 q_1 p_2 + \epsilon_1 \epsilon_2 p_2 q_1 q_1 p_1 q_2
+ \epsilon_1 p_1 q_2 q_1 p_2 q_1~.
\eqno\rm(8)
$$
The three terms result from the three possible $q_1$ factors to
differentiate, and the $\epsilon$'s come from
cyclically permuting the factors to
bring the particular $q_1$ which is to be differentiated to the right.

The first term on the right hand side of Eq.~(6), expanded out in this
notation, is
$$ {\bf\{A,\{B,C\}\}} = \{{\bf A},\epsilon_r
\left( {\bf B}_r {\bf C}^r -
                 {\bf C}_r {\bf B}^r \right)\}~,
\eqno\rm (9a)
$$
which can be expanded further to
$$
{\bf\{A,\{B,C\}\}} = \epsilon_r\epsilon_s \left(
{\bf A}_s \left({\bf B}_r{\bf C}^r\right)^s-
{\bf A}_s\left({\bf C}_r{\bf B}^r\right)^s-
\left({\bf B}_r{\bf C}^r\right)_s{\bf A}^s
+\left({\bf C}_r{\bf B}^r\right)_s{\bf A}^s
\right)~.
\eqno\rm (9b)
$$
Cyclic permutations of ${\bf A, B}$, and ${\bf C}$ give the other two terms
in Eq.~(6).  Thus, the left--hand side of Eq.~(6) is
$$
{\bf [A,B,C]}=\epsilon_r\epsilon_s[\left(
{\bf A}_s \left({\bf B}_r{\bf C}^r\right)^s
-{\bf A}_s\left({\bf C}_r{\bf B}^r\right)^s-
\left({\bf B}_r{\bf C}^r\right)_s{\bf A}^s
+\left({\bf C}_r{\bf B}^r\right)_s{\bf A}^s
\right)
$$
$$
\quad\quad\quad\quad \quad\quad +\left(
{\bf B}_s \left({\bf C}_r{\bf A}^r\right)^s
-{\bf B}_s\left({\bf A}_r{\bf C}^r\right)^s-
\left({\bf C}_r{\bf A}^r\right)_s{\bf B}^s
+\left({\bf A}_r{\bf C}^r\right)_s{\bf B}^s
\right)
\eqno\rm(10)
$$
$$
\quad\quad\quad\quad \quad\quad +\left(
{\bf C}_s \left({\bf A}_r{\bf B}^r\right)^s
-{\bf C}_s\left({\bf B}_r{\bf A}^r\right)^s-
\left({\bf A}_r{\bf B}^r\right)_s{\bf C}^s
+\left({\bf B}_r{\bf A}^r\right)_s{\bf C}^s
\right)]~.
$$

Let us first consider how the terms in Eq.~(10) cancel in the classical,
c--number case.  A similar cancellation mechanism will also apply in the more
general quantum operator case.  For c--numbers, the trace operation is trivial,
derivatives of functionals commute, and one can apply the product rule
to expand the terms.  For instance, $$
\left({\bf B}_r{\bf C}^r\right)^s = {\bf B}_r^{~s}{\bf C}^r+
{\bf B}_r{\bf C}^{rs}~.
\eqno\rm(11)
$$
Note that ${\bf B}_r^{~s}$ means that the $q_r$ derivative is applied before
the
$p_s$ derivative.  ${\bf B}_{~r}^s$ would mean that the same derivatives
are applied in the opposite order.  This distinction is
meaningless for c--number fields, where derivatives commute, but it is
crucial for non--commutative operators $\{q_r\}$ and $\{p_r\}$.

Equation~(11) implies that each term in Eq.~(10) will generate two terms.
These
terms cancel in pairs.  For example, in the first term in Eq.~(10),
consider the derivative with respect to $p_s$ applied to ${\bf B}_r$.
This generates the term $+{\bf A}_s{\bf B}_r^{~s}{\bf C}^r$.  This
cancels against the term $-{\bf A}_r{\bf B}_{~s}^{r}{\bf C}^s$
obtained by applying the derivative with respect to $p_s$ on ${\bf B}_r$ in
the eleventh term (the dummy indices $r$ and $s$ need to be interchanged
for the terms
to be the same).  The other half of the eleventh term will in turn be
cancelled by a part of the eighth term, and so on.  After twelve such
double terms have been computed, we come back to the beginning and all
terms have been cancelled.

The order in which these cancellations occur is as
follows,
$$
\eqalign{
\longleftrightarrow
({\bf A}_s \left( {\bf B}_r{\bf C}^r \right)^s)
\longleftrightarrow
(\left( {\bf A}_s{\bf B}^s \right)_r {\bf C}^r)
\longleftrightarrow
(\left( {\bf A}_r{\bf C}^r \right)_s {\bf B}^s)
\longleftrightarrow
({\bf A}_r \left( {\bf C}_s{\bf B}^s \right)^r)
\longleftrightarrow
\cr ({\bf C}_s \left( {\bf A}_r{\bf B}^r \right)^s)
\longleftrightarrow
(\left( {\bf C}_s{\bf A}^s \right)_r {\bf B}^r)
\longleftrightarrow
(\left( {\bf C}_r{\bf B}^r \right)_s {\bf A}^s)
\longleftrightarrow
({\bf C}_r \left( {\bf B}_s{\bf A}^s \right)^r)
\longleftrightarrow
\cr ({\bf B}_s \left( {\bf C}_r{\bf A}^r \right)^s)
\longleftrightarrow
(\left( {\bf B}_s{\bf C}^s \right)_r {\bf A}^r)
\longleftrightarrow
(\left( {\bf B}_r{\bf A}^r \right)_s {\bf C}^s)
\longleftrightarrow
({\bf B}_r \left( {\bf A}_s{\bf C}^s \right)^r)
\longleftrightarrow
}
\eqno\rm(12)
$$
where we have used the fact that $r$ and $s$ are dummy indices and
have interchanged them in some of the terms, and where the lower right
of Eq.~(12) links back to the upper left.  By Eq.~(11), each entry in
Eq.~(12) generates two terms; one of these
cancels against a term from the entry to the immediate left in the
chain, and the other cancels against a term from the entry to the
immediate right.

We will now proceed to show that in the general case, the
cancellations occur in a similar way.  However, the absence of both
commutativity and the product rule makes the proof a little less
trivial.  For the rest of this discussion, we will also not assume the
summation convention, which means that repeated indices are {\it not}
summed henceforth, so that we are dealing with summands which appear,
summed over $r$ and $s$, in the Jacobi identity.  Also, we will assume
that ${\bf A}$, ${\bf B}$,
${\bf C}$ are monomials in $\{q_r\}$ and $\{p_r\}$.  The proof for the general
case of polynomial functionals follows from expanding out the
generalized Poisson brackets in Eq.~(6) in terms of monomials.  Finally,
recall that there is an implied trace ${\bf Tr}$ for each
pair of parentheses $(~)$ in Eq.~(10).
This means
that we can cyclically permute the factors within a parenthesis, if we
include a factor
$\epsilon_r$ every time a $q_r$ or $p_r$ is moved from the front of
a trace to the back.  Thus, in our shorthand notation, $(q_r{\cal O})$ =
$\epsilon_r ({\cal O}q_r)$.

When one computes the derivative of some monomial
with respect to $q_r$ (say), each particular occurance of $q_r$ generates
one term in the result.  Consider the expression,
$$
\left({\bf B}_r{\bf C}^r\right)^s~,
\eqno\rm(13)
$$
which appears in the first term of Eq.~(12).  In this
expression, there are three derivatives, and a choice is made of which
occurance of $q_r$, $p_r$, and $p_s$ to differentiate in the
appropriate terms.  Each set of choices will produce a particular
monomial term in the result.  If $q_r$ appears $N({\bf B},q_r)$ times
in the monomial ${\bf B}$, and $p_r$ appears $N({\bf C},p_r)$ times in C,
and so on, then the number of terms produced by Eq.~(13) is at most $N({\bf
B},q_r)N({\bf C},p_r)(N({\bf B},p_s)+N({\bf C},p_s))$.

We will show that in Eq.~(10), each such monomial term in the result, for
fixed $r$,
$s$ ({\it i.e.}, for a fixed choice of $q_r, p_r, q_s, p_s$), will cancel
with its counterpart in the order defined by Eq.~(12).  Consider the
case where the $p_s$ derivative is applied to ${\bf B}$
in the first entry and the $q_r$ derivative is applied to ${\bf B}$ in the
second entry of Eq.~(12).  For these to give non--vanishing
contributions, ${\bf B}$
must contain at least one instance of both $q_r$ and $p_s$.  Therefore
the most general form for ${\bf B}$ is
$$
{\bf B} = (\alpha q_r\beta p_s)~,
\eqno\rm(14)
$$ where $\alpha$ and $\beta$ are arbitrary monomials (and could
possibly contain $q_r$ and $p_s$).
The displayed $q_r$ and $p_s$
are the particular instances of these coordinates in ${\bf B}$ upon which
the derivatives will act.

We have
$$
\eqalignno{({\bf A}_s({\bf B}_r{\bf C}^r)^s)
&=({\bf A}_s((\alpha q_r\beta p_s)_r{\bf C}^r)^s)\cr
&= \epsilon_{\alpha}\epsilon_r ({\bf A}_s(\beta p_s\alpha {\bf
C}^r)^s)\cr
&= \epsilon_{\alpha}\epsilon_r \epsilon_{\beta} \epsilon_s
({\bf A}_s\alpha {\bf C}^r\beta)~,&(15)\cr}
$$
and
$$
\eqalignno{(({\bf A}_s{\bf B}^s)_r{\bf C}^r)
&=(({\bf A}_s(\alpha q_r\beta p_s)^s)_r {\bf C}^r)\cr
&= (({\bf A}_s\alpha q_r \beta)_r{\bf C}^r )\cr
&= \epsilon_{\beta}(\beta {\bf A}_s \alpha {\bf C}^r)\cr
&=({\bf A}_s\alpha {\bf C}^r\beta)~. &(16)\cr}
$$

If ${\bf B}$ is not identically zero, making the equality of Eqs.~(15) and
(16) trivial, it
must have an even number of fermion factors.  Therefore,
$\epsilon_{\alpha}
\epsilon_r\epsilon_{\beta}\epsilon_s=1$,
and so the right hand sides of Eqs.~(15) and (16) are always the same.
Finally, these same cancellations can be
shown to occur for every summand term in Eq.~(10) in the order indicated by
Eq.~(12), and apply both to the summands with $r\not= s$ and to those
with $r=s$, including the parts of the summands with $r=s$ in which
there are two derivatives with respect to the same variable $q_r$ (or
$p_r$).  This proves that the
Jacobi identity is true for arbitrary bosonic and fermionic quantum field
operator variables $\{q_r\}$ and $\{p_r\}$.
\bigskip
\centerline{\bf Acknowledgements}
This work was supported in part by the
Department of Energy under Grant \#DE--FG02--90ER40542.  GVB and JDW
thank Brookhaven National Laboratory for its hospitality while this
work was done.  SLA acknowledges the hospitality of the Aspen Center for
Physics, and thanks Freeman Dyson and Yong--Shi Wu for
conversations about the Jacobi identity.

\vfill\eject
\centerline{\bf References}
\Rf
[1]  S.L. Adler, Generalized quantum dynamics, Institute for Advanced Study
preprint, IASSNS-HEP-93/32, June, 1993.
\Rf
[2]  S.L. Adler, Quaternionic quantum mechanics and quantum fields
(Oxford University Press, Oxford, 1994).
\Rf
[3]  E. Witten, J. Diff. Geom. 17 (1982) 661.
\bye